# Countermeasure against quantum hacking using detection statistics

Gaëtan Gras,[1,2] Davide Rusca,[2] Hugo Zbinden,[2] and Félix Bussières[1,2]
[1]*ID Quantique SA, CH-1227 Carouge, Switzerland*
[2]*Group of Applied Physics, University of Geneva, CH-1211 Geneva, Switzerland*

Detector blinding attacks have been proposed in the last few years, and they could potentially threaten the security of QKD systems. Even though no complete QKD system has been hacked yet, it is nevertheless important to consider countermeasures to avoid information leakage. In this paper, we present a new countermeasure against these kind of attacks based on the use of multi-pixel detectors. We show that with this method, we are able to estimate an upper bound on the information an eavesdropper could have on the key exchanged. Finally, we test a multi-pixel detector based on SNSPDs to show it can fulfill all the requirement for our countermeasure to be effective.

## I. INTRODUCTION

Since its first proposal by Bennett and Brassard in 1984 [1], quantum key distribution (QKD) has attracted a lot of interest for securing communications. Indeed, with QKD, two distant parties, Alice and Bob can securely exchange a key to encrypt their communications. QKD does not require making assumptions on the computational power of the eavesdropper Eve, making this technology theoretically secure. However, imperfections of physical systems can potentially be exploited by Eve to break the security and obtain some information on the key without being noticed. Several attacks have already been proposed such as the photon-number splitting (PNS) attack [2], detector efficiency mismatch attack [3], Trojan horse attack [4–6] as well as potential countermeasures such as use of decoy states [7–9] to estimate the amount of information shared with Eve.

In this paper, we are interested in detector control attacks such as blinding attacks [10–13]. When no countermeasure is in place, this attack could possibly allow Eve to gain full information on the key exchange by Alice and Bob without being noticed. Some protocols such as device-independent (DI) [14–18] or measurement-device-independent (MDI) [19–24] are secure against these attacks but their current performances and certain technical challenges could hamper their deployment of a large scale QKD network in the near future. For other protocols, like prepare-and-measure (PM) protocols, several potential countermeasures have been proposed like monitoring the state of the detector [25, 26], using a variable optical attenuator [27–29] or with specially design readout circuit [30, 31]. Here, we propose a novel method solely based on detection statistics using multi-pixel detectors to estimate the maximum information Eve can have on the key exchanged.

In the first section, we detail the scheme of the attack considered and we present the security principle of our countermeasure using a simple case. Then, we give results of our analysis in more realistic conditions. Finally, we test a 2-pixels detector under blinding attack and show that it can fulfill the requirements for our countermeasure.

## II. COUNTERMEASURE

Blinding attacks have been shown to potentially threaten the security of QKD. Indeed, it gives the possibility to an adversary Eve to change the behavior of Bob's detectors such that she can send what is usually called a "faked state" that can only be detected if Bob's chooses the same basis as hers. In this way, Eve can reproduce her measurement outcome without introducing errors in the key. As a countermeasure, we propose to split Bob's detectors into two pixels. Other implementations could be possible but we show in Sec. III that the 2-pixel detector is a good way to to it. As both pixels correspond to the detection of the same state, our main assumption is that Eve's faked state cannot be used to control each pixel independently, and that the coincidence detection probability in the presence of the faked states will inevitably rise, and reveal Eve's attack. More precisely, we show that the measurement of the probabilities of single and coincidence detections gives enough information to Alice and Bob to estimate the maximum amount of information an eavesdropper can have on the key.

The scheme of the attack is shown in Fig. 1. Alice sends weak coherent pulses (WCP) with a mean photon number $\mu$. Bob's measurement setup is composed of a basis choice (active or passive) and two detectors each split into two pixels. Eve is in the middle and can either perform the blinding attack or simply let the pulse from Alice go through to Bob. We note $p_a$ the probability of attack. If Eve lets Alice's pulse go through, Bob's pixel $i \in \{1,2\}$ will click with a probability $p_{B1} = (1+\alpha)p_B$ and $p_{B2} = (1-\alpha)p_B$, where $p_B$ is the average pixel detection probability and $\alpha$ is a coefficient known by Bob characterizing the efficiency mismatch between the pixels. If Eve chooses to intercept Alice's pulse, she measures it using a copy of Bob's setup (called 'fake Bob') and she resends her faked state if she detected something. Bob's pixel $i$ will detect this faked state with a probability $p_{di}$ only if his basis choice is the same as Eve's. Otherwise, he will not detect anything. Therefore, the detection probability when Eve does her attack depends on the probability that Alice's pulse contains at least one photon $1 - \mathrm{e}^{-\mu t}$ ($t$ being the transmission coefficient between Alice and Eve's detectors) and on the probability

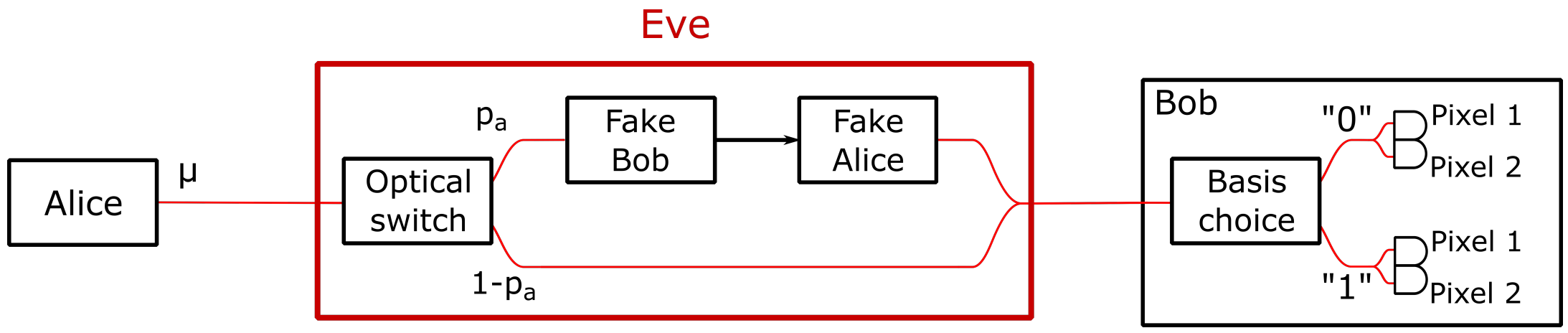

FIG. 1. Scheme of the attack. Alice sends pulses with a mean photon number per pulse $\mu$. Eve intercepts the pulse with a probability $p_a$. If she gets a conclusive event with her 'fake Bob', she resends a pulse to force Bob's detector to click; otherwise, she does nothing.

$q$ that Bob and Eve choose the same basis. We call this probability $p_E$ :

$$p_E = (1 - \mathrm{e}^{-\mu t})q. \tag{1}$$

By naming $p_{s1}$ and $p_{s2}$ the probabilities of detection of both pixels measured by Bob, we then can write :

$$\begin{aligned} p_{s1} &= p_a p_E \sum_\lambda p^\lambda p_{d1}^\lambda + (1-p_a)(1+\alpha)p_B, \\ p_{s2} &= p_a p_E \sum_\lambda p^\lambda p_{d2}^\lambda + (1-p_a)(1-\alpha)p_B. \end{aligned} \tag{2}$$

We give Eve the possibility to use different strategies $\lambda$ from one pulse to the other, each with a probability $p^\lambda$. We suppose both pixels are independent from each other. Therefore, the probability that a faked state generates a coincidence is $p_{d1}p_{d2}$. The probability of coincidence for the two pixels is then :

$$p_c = p_a p_E \sum_\lambda p^\lambda p_{d1}^\lambda p_{d2}^\lambda + (1-p_a)(1-\alpha^2)p_B^2. \tag{3}$$

By analyzing the coincidence probability between both pixels, we show how Alice and Bob can bound the information leaked to Eve.

### A. Asymptotic case

In this section, we first want to convey the idea behind this countermeasure by considering a simple case where we are in the asymptotic limit and both pixels are perfectly identical ($p_{d1}^\lambda = p_{d2}^\lambda$ and $p_{B1} = p_{B2}$). The attack scenario defined by Eqs. (2) and (3) can be rewritten :

$$\begin{aligned} p_s &= p_a p_E \sum_\lambda p^\lambda p_d^\lambda + (1-p_a)p_B, \\ p_c &= p_a p_E \sum_\lambda p^\lambda (p_d^\lambda)^2 + (1-p_a)p_B^2. \end{aligned} \tag{4}$$

We define the ratio $r = p_c/p_s^2$ (note, this is similar to a second-order correlation measurement $g_2$; we call it $r$ simply because with the attack, it is not really a measurement on the photon statistics). In the limit $p_a = 0$, $r = 1$ as expected for coherent states. On the other hand, if $p_a = 1$, we have :

$$r = \frac{p_c}{p_s^2} = \frac{\sum_\lambda p^\lambda (p_d^\lambda)^2}{p_E \left(\sum_\lambda p^\lambda p_d^\lambda\right)^2} \geq \frac{1}{p_E} > 1. \tag{5}$$

As we can see, the value of $r$ induced by the attack is limited by the probability $p_E$ which depends on the vacuum probability in Alice's pulses and $q$. Let's now see how we can estimate Eve's information per bit $I_E$ on the raw key in the case she attacks only a fraction of the pulses i.e. $0 < p_a < 1$. As Eve knows the measurement outcome of Bob only when he detects a faked state, we want to maximise :

$$I_E = \frac{p_a p_E \sum_\lambda p^\lambda p_d^\lambda}{p_s}, \tag{6}$$

given $p_E$, $p_s$ and $p_c$. Using the Lagrangian multiplier, we can show that Eve's best strategy is to always resend a pulse with the same probability of detection $p_d^\lambda = p_d$, $\forall\lambda$ and we find her maximum information is given by (see Appendix A 1):

$$I_{E,max} = \frac{\sqrt{p_E}(\sqrt{p_c} - p_s)}{p_s(1-\sqrt{p_E})} = \frac{\sqrt{p_E}}{(1-\sqrt{p_E})}\left(\sqrt{r} - 1\right). \tag{7}$$

Figure 2 shows the values of $I_{E,max}$ as a function of $r$. As expected, Eve's information is linked to the ratio $r = p_c/p_s^2$ measured by Bob.

In order to make our approach more realistic, we have to consider that the Bob's pixels are not perfectly identical. Their quantum efficiency may not be the same and/or their response to the attack is certainly different, even slightly. We come back then to Eqs. (2) and (3). If we do not put any constraint on $p_{d1}^\lambda$ and $p_{d2}^\lambda$, we give Eve all the power possible making our countermeasure inefficient. On the other hand, a complete characterization of all detectors under all possible attack conditions in order to find bounds on $p_{d2}$ given $p_{d1}$ seems an unpractical task. We circumvent this problem by adding the assumption that one pixel will always detect Eve's

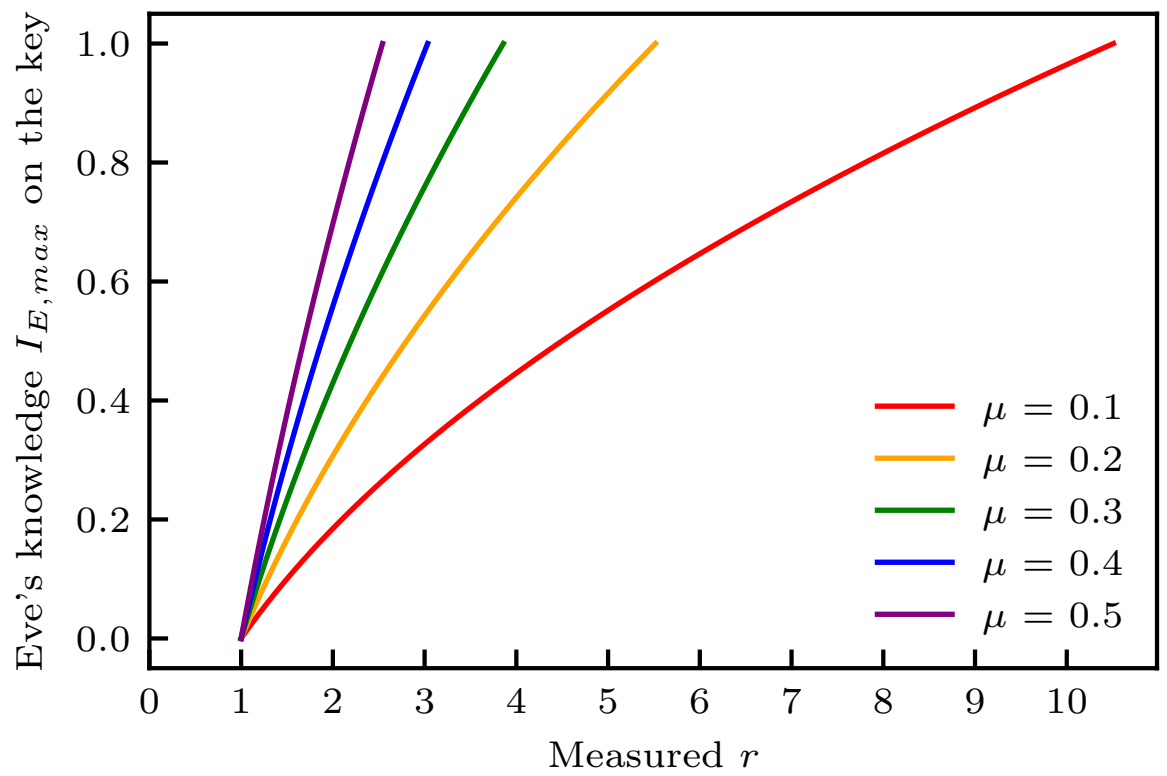

FIG. 2. Eve's information on the raw key exchange by Alice and Bob versus the ratio $r$ measured by Bob.

faked state with a higher probability. This constraint on the attack can be written as

$$\text{sign}(p_{d1}^{\lambda} - p_{d2}^{\lambda}) \neq f(\lambda). \tag{8}$$

We show in Sec. III that this condition can be realized with a two-pixel detector. By applying the Lagrange multiplier, we can calculate all the extrema of $I_E$ to find the maximum of Eve's information $I_{E,max}$. Here, we limit the number of strategies to 2 as increasing the number of strategies does not give much more information to Eve if the difference between $p_{s1}$ and $p_{s2}$ stays small. Indeed, in that case, Eve is forced to make both pixels click with the same probability most of the time to keep the probabilities of detection close. In a real system, the protocol can be aborted if the difference between $p_{s1}$ and $p_{s2}$ exceeds a certain threshold. Details of the calculations are given in Appendix A 2.

### B. Finite key analysis

In order to take into account finite key length effects, we need to bound the probabilities of single and coincidence measured by Bob. Usually, QKD proofs rely on Hoeffding's inequality to calculate upper and lower bounds on measured values. However, in our countermeasure, the probability of coincidence will drop very quickly with the quantum channel length and in this case, Hoeffding's inequality is no longer tight. This would lead to an overestimation of Eve's information making our countermeasure usable only for short distances. In order to have a tighter bound on Bob's probabilities, we can use the equations given in [32]. The upper and lower bounds on $p_{si}$ and $p_c$ are given by :

$$\begin{aligned} p_c^u &= 1 - I_{\epsilon}^{-1}(N(1-p_c), Np_c + 1), \\ p_{si}^l &= I_{\epsilon}^{-1}(Np_{si}, N(1-p_{si}) + 1), \end{aligned} \tag{9}$$

where $N$ is the total number of pulses sent by Alice, $\epsilon$ our confidence factor and $I^{-1}$ the inverse incomplete beta function. By inserting these bounds in the calculation of $I_{E,max}$, we obtain an upper bound on Eve's information $I_{E,max}^u$.

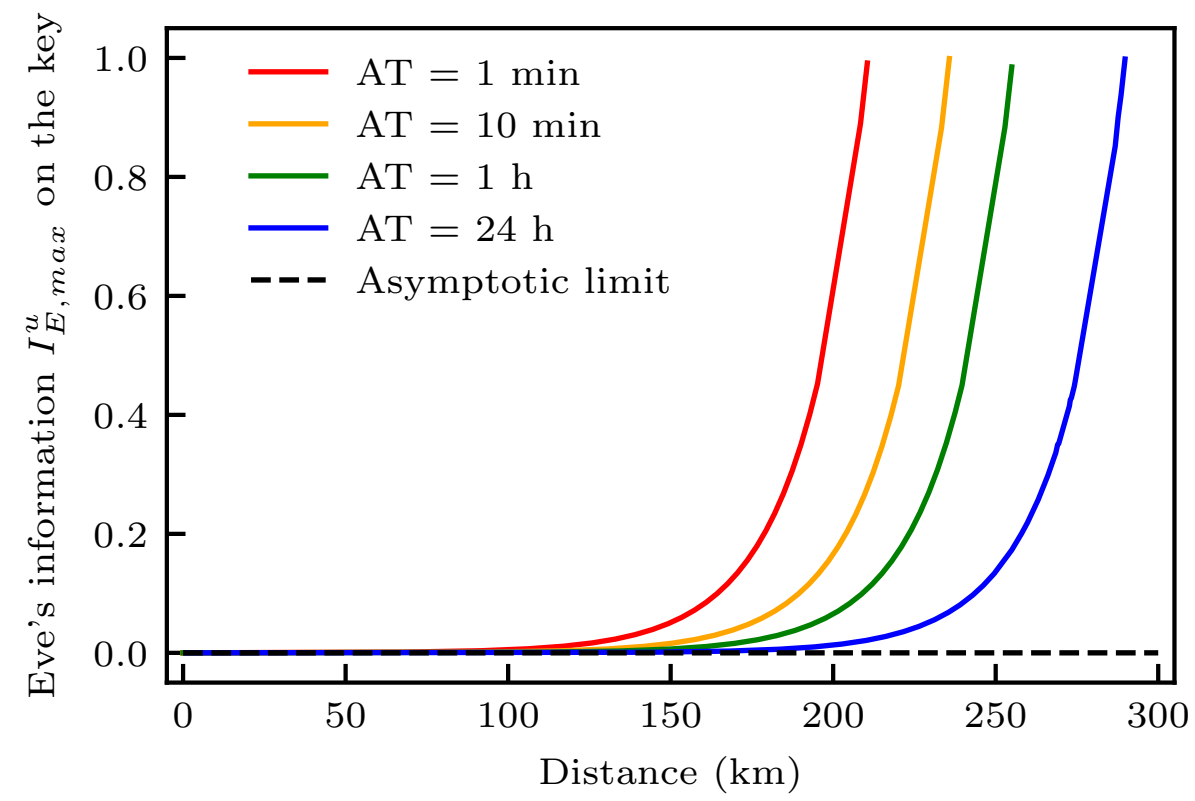

FIG. 3. Upper bound on Eve's information of the raw key as a function of the channel length between Alice and Bob for different acquisition times (AT). The protocol used is a BB84 with an active basis choice. Alice sends pulses with a mean photon number $\mu = 0.5$ at a rate of 10 GHz. Losses in the channel are 0.2 dB/km. Bob's pixels have a quantum efficiency of 50%.

Figure 3 shows simulations of $I_{E,max}^u$ for a BB84 protocol. We run the simulations for different acquisition time for Bob. As the quantum channel length increases, the probability of coincidence measured by Bob decreases rapidly requiring longer acquisition times to limit the uncertainty. Therefore, the factor ultimately limiting our countermeasure is the acquisition time allowed by Alice and Bob. Nevertheless, acquisition times of less than 24 hours are sufficient for our countermeasure to be working for distances of almost 250 km which is close to the limit of many current QKD implementations.

## III. EXPERIMENTAL RESULTS

In this section, we show that actual detectors can fulfill the condition given by Eq. (8) for our countermeasure against blinding attacks. To do so, we fabricated and tested multi-pixel superconducting nanowire single-photon detectors (SNSPDs), as depicted on Fig. 4a, as attacks on this kind of detector have been reported [12, 13, 33]. With this design, both pixels are illuminated by a single fiber limiting the dependency of the light distribution on the wavelength used by Eve for her attack compared to an implementation with a beam splitter and two distinct detectors. For even better security, the addition of a mode scrambler could prevent Eve from using smaller wavelength where the fiber becomes multimode [34].

To illustrate how the blinding attack on a QKD system using this kind of detector works, we take as an example a BB84 protocol in polarization. In normal operation, when a photon hits the SNSPD, it will break the superconductivity inducing a rapid increase of the resistance of the nanowire. This sudden change of resistance will divert the bias current of the detector toward the readout circuit to generate a click. In order to blind Bob's detectors, Eve sends unpolarized light of a few hundreds of nW inside Bob's setup such that her blinding power is equally distributed over all detectors. This forces the SNSPDs to stay in a resistive state where they are insensitive to single photon. When Eve wants to force Bob to detect the state of her choice, say $|H\rangle$, she polarizes her blinding light vertically for a time $\Delta t$. During this time, the optical power arriving on detector $D_H$ will be greatly reduced (around 20 to 30 dB depending on Bob's components) while keeping the other detectors blinded.

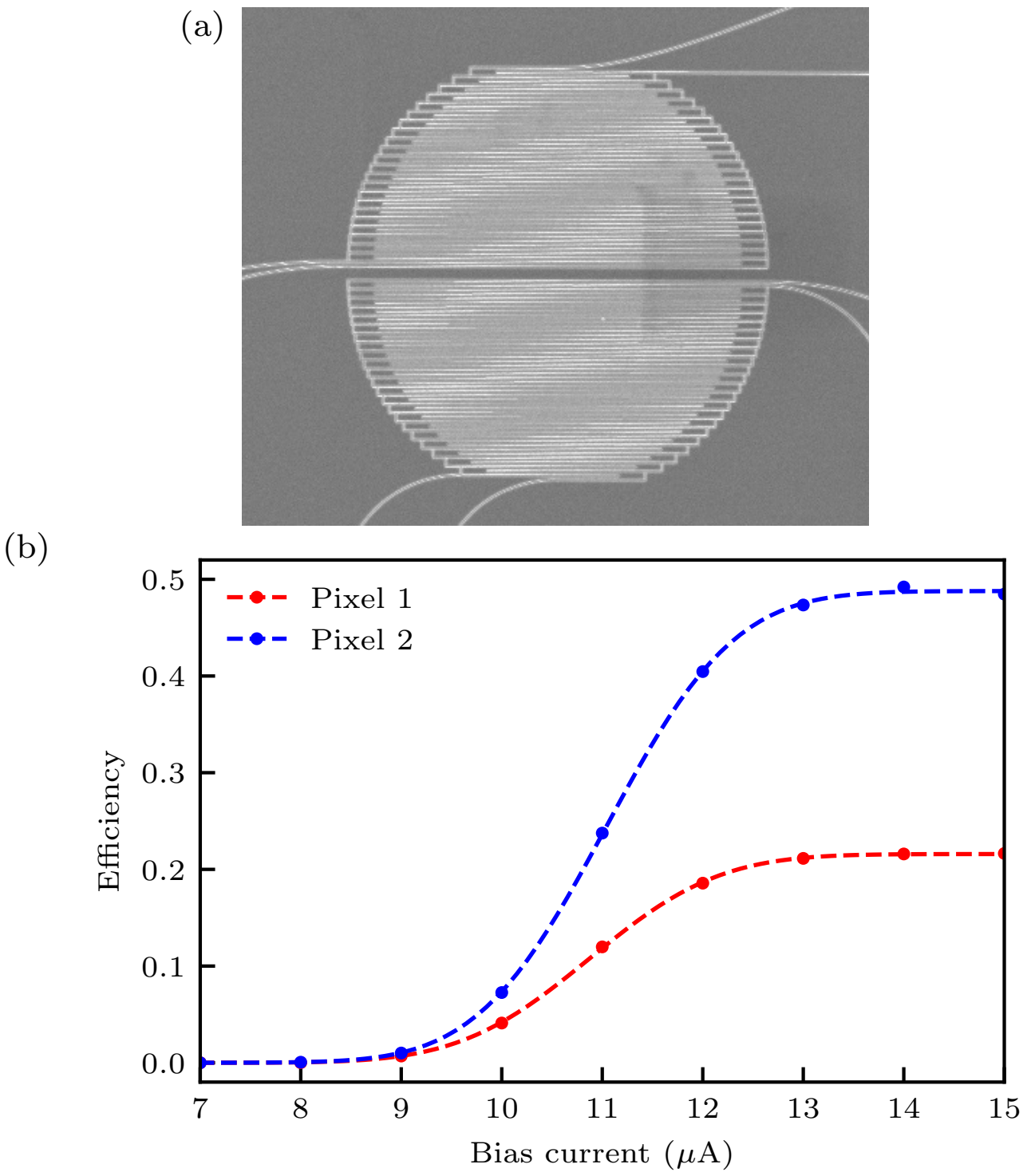

FIG. 4. (a) SEM image of a two-element molybdenum silicide (MoSi) superconducting nanowire single-photon detector (SNSPD). Each pixel has its own bias current and readout circuit. The nanowire width is 100 nm with a fill factor of 0.6 [35]. The two pixels are separated by 600 nm to avoid thermal-crosstalk between them. (b) Efficiency curves at 1550 nm of the two pixels of the detector operated at 0.8 K versus the bias current.

By unpolarizing her blinding light after $\Delta t$, the optical power $P_H$ arriving on the detector $D_H$ will increase suddenly, forcing it to click as part of the current would have returned to the nanowire (see Fig. 5). Eve can control the probability $p$ to force the detector to click by allowing more or less current to return to the detector via $\Delta t$. Many parameters have an influence on the probability of detection of the faked state. Some are controlled by Eve (blinding power $P_{blind}$, $\Delta t$) and some are controlled by Bob (bias current). However, as we mentioned in Sec. II A, if we can find a regime where one pixel always clicks with a probability greater than the second one (whatever are the parameters of the attack) then this gives enough constraints on Eve to ensure she cannot steal the key without being noticed. As the probability of click depends on the amount of current that returned to the nanowire, we want one pixel to recover its current more rapidly such that it will detect the faked state with a higher probability than the second pixel. For that, we set pixel 2 at its maximum bias current (15 µA) while pixel 1 is set at a bias current of 12.5 µA. This way, the current will return more rapidly to the pixel 2 [36].

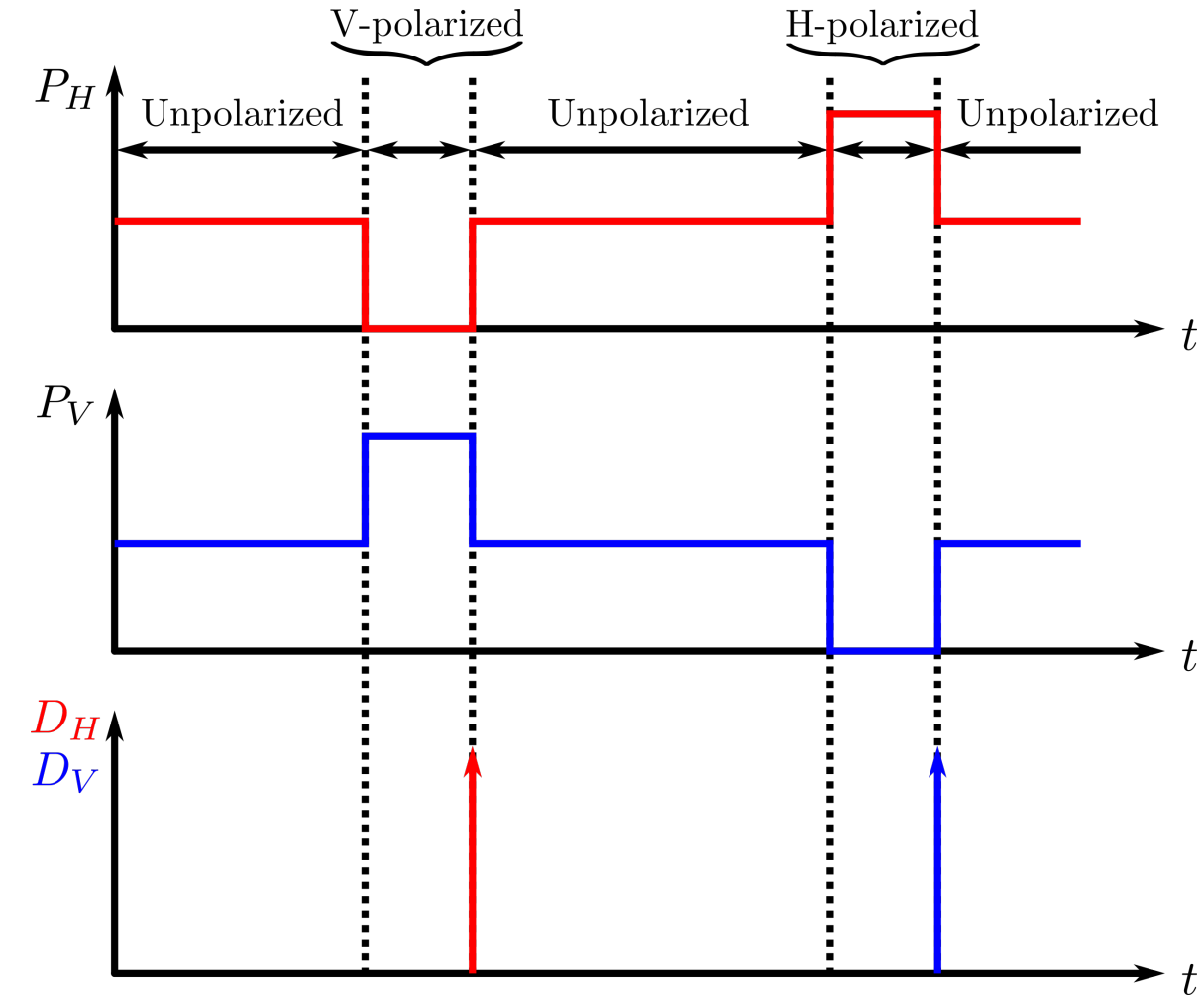

FIG. 5. Schematic representation of the blinding power distribution on detectors $D_H$ and $D_V$ during the attack on a BB84 QKD protocol based on polarization. By changing the polarization of her blinding light, Eve can let the detector of her choice partially recover its bias current to force it then to click.

In that configuration, we can see in Fig. 6a that $p_{d2} > p_{d1}$ for the all range of working $P_{blind}$ and $\Delta t$ without impacting the efficiency of the two pixels. The limitation of this configuration will only appear in high rate QKD protocols were the time dynamics of the detectors begins to be impacting. We also compared the coincidence probability due to the faked state with the product of the detection probability of each pixel in order to verify that both pixels are independent under the blinding attack. Results are shown in Fig. 6b. As we can see, both values are in the uncertainty range of each over, validating the assumption made in our analysis. Thus, this multi-pixel detector fulfills all the requirements for our countermeasure.

This countermeasure could also work with single-photon avalanche diodes (SPAD) detectors as the core idea behind our proposal does not rely on the working principle of the detectors. Further tests with this kind of detectors needs to be done to validate that it fulfills all the necessary conditions.

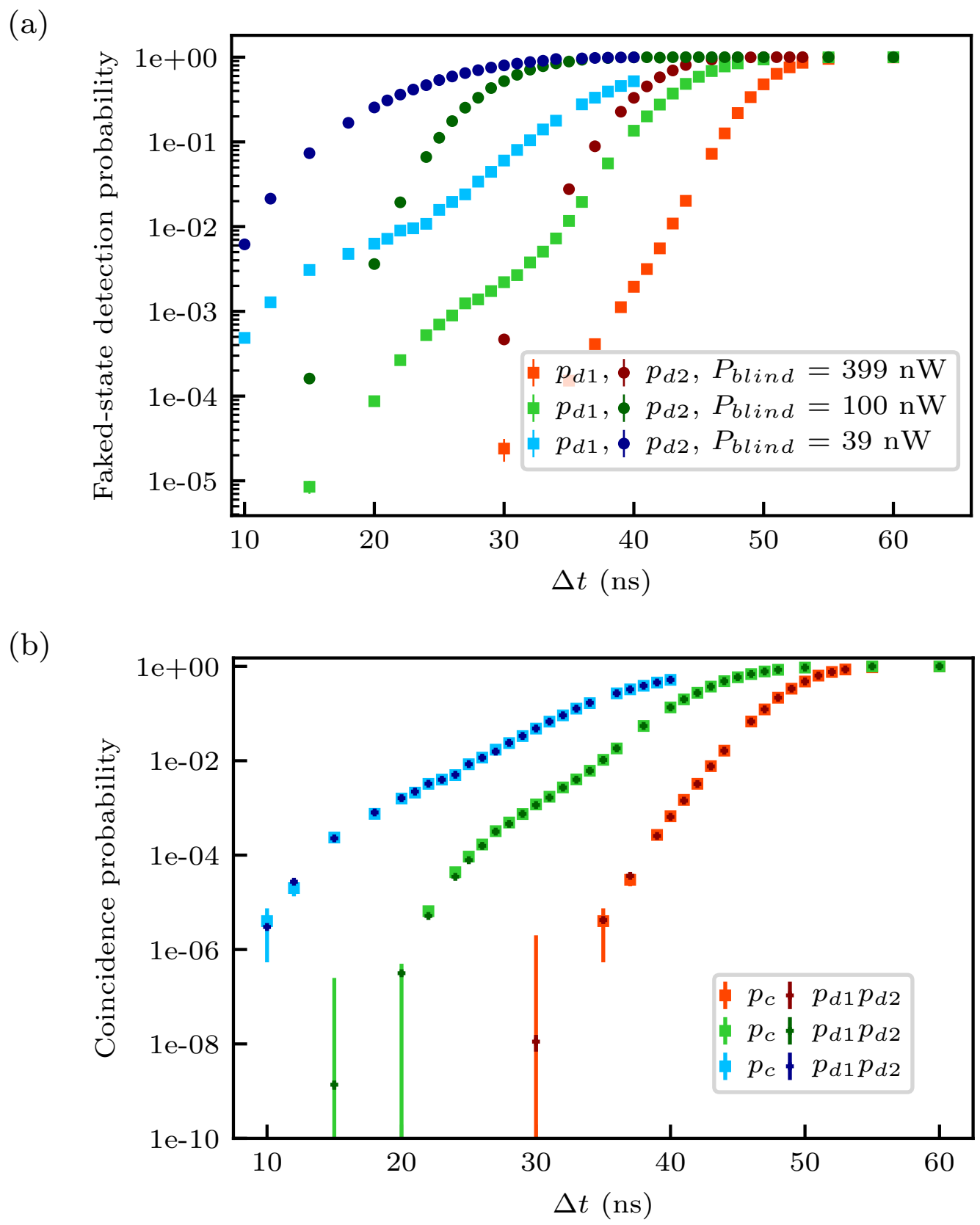

FIG. 6. (a) Probability of detection of the faked state. Pixel 1 : $I_{b1} = 12.5$ µA; pixel 2 : $I_{b2} = 15$ µA. We varied the blinding power between 39 nW and 399 nW as it was the working range for the blinding attack. (b) Comparison between the measured coincidence probability and the coincidence probability calculated from the faked-state detection probabilities of both pixels.

## IV. CONCLUSION

In this paper, we proposed a countermeasure against detector control attack based on multi-pixel detectors. With this method, we are able to estimate an upper bound on the information leaked to the adversary solely using the single and coincidence probabilities measured by Bob. The effectiveness of our countermeasure over long distances is ultimately limited by the key exchange time between Alice and Bob. Nevertheless, we showed that communications close to 250 km can be secured against the attack with acquisition time of less than 24 hours. Finally, we experimentally demonstrated that a multi-pixel SNSPD operated in the right conditions by Bob can satisfy the assumptions made in our analysis.

## ACKNOWLEDGMENTS

This project has received funding from the European Union's Horizon 2020 research and innovation programme under the Marie Skłodowska-Curie grant agreement N° 675662. We thank Claire Autebert for designing and fabricating the detectors. We also thank Jean-Daniel Bancal for helpful discussions.

## Appendix A: Lagrange multiplier calculations

### 1. Simple case

In order to find Eve's best strategy, we want to maximise the number of detections coming from faked states $n_a = N p_a p_E \sum_\lambda p^\lambda p_d^\lambda$ (with $n$ being the total number of pulses sent by Alice) over the total number of detections $n$ under the constraints given by Eq. (4). As $n$ and $N$ are fixed value, we can maximize the function $f$ defined by :

$$f = p_a p_E \sum_\lambda p^\lambda p_d^\lambda. \tag{A1}$$

We define the following equations representing our constraints :

$$\begin{aligned} g_1 &= p_a p_E \sum_\lambda p^\lambda p_d^\lambda + (1-p_a)p_B - p_s, \\ g_2 &= p_a p_E \sum_\lambda p^\lambda (p_d^\lambda)^2 + (1-p_a)p_B^2 - p_c, \\ g_3 &= \sum_\lambda p^\lambda - 1. \end{aligned} \tag{A2}$$

We can then define our Lagrange function :

$$\mathcal{L}(p_a, p^\lambda, p_d^\lambda, p_B, \Lambda_1, \Lambda_2, \Lambda_3) = f - \Lambda_1 g_1 - \Lambda_2 g_2 - \Lambda_3 g_3. \tag{A3}$$

The function $f$ is maximum if :

$$\nabla \mathcal{L} = 0. \tag{A4}$$

To show that Eve's best strategy is to always always send faked state with the same probability of detection, we take the derivatives :

$$\begin{aligned} \frac{\partial \mathcal{L}}{\partial p_d^\lambda} &= p_a p_E p^\lambda - \Lambda_1 p_a p_E p^\lambda - 2\Lambda_2 p_a p_E p^\lambda p_d^\lambda \\ &= p_a p_E p^\lambda \left(1 - \Lambda_1 - 2\Lambda_2 p_d^\lambda\right) \\ &= 0. \end{aligned} \tag{A5}$$

This expression is valid only if $1 - \Lambda_1 - 2\Lambda_2 p_d^\lambda = 0,\ \forall\lambda$ (we neglect the case $p_a = 0$ as it would mean that Eve never does the attack and the case $p^\lambda = 0$ as it would be a strategy Eve never uses). Therefore, either $p_d^\lambda$ is a constant or $\Lambda_1 = 1$ and $\Lambda_2 = 0$. The latter case is impossible as we can see by looking at an another derivative :

$$\begin{aligned} \frac{\partial \mathcal{L}}{\partial p_B} &= -(1-p_a)(\Lambda_1 + 2\Lambda_2 p_B) \\ &= 0. \end{aligned} \tag{A6}$$

The solution $p_a = 1$ is possible only if $p_c/p_s^2 \geq 1/p_E$. Otherwise, $\Lambda_1 + 2\Lambda_2 p_B = 0$ which is incompatible with $(\Lambda_1, \Lambda_2) = (1, 0)$. Consequently, Eve's best strategy is to use the same $p_d^\lambda = p_d$, $\forall\lambda$. These results simplify our problem that we can rewrite as follow :

$$\begin{aligned} f &= p_a p_E p_d, \\ g_1 &= p_a p_E p_d + (1-p_a)p_B - p_s, \\ g_2 &= p_a p_E p_d^2 + (1-p_a)p_B^2 - p_c, \\ \mathcal{L} &= f - \Lambda_1 g_1 - \Lambda_2 g_2, \\ \nabla\mathcal{L} &= 0. \end{aligned} \tag{A7}$$

This system has a unique solution :

$$\begin{aligned} p_B &= \sqrt{p_c}, \\ p_d &= \sqrt{\frac{p_c}{p_E}}, \\ p_a &= \frac{\sqrt{p_c} - p_s}{\sqrt{p_c}(1-\sqrt{p_E})}, \end{aligned} \tag{A8}$$

which finally gives us:

$$\begin{aligned} I_{E,max} &= \frac{n_a}{n} \\ &= \frac{\sqrt{p_E}(\sqrt{p_c} - p_s)}{p_s(1-\sqrt{p_E})}. \end{aligned} \tag{A9}$$

### 2. General case

In the general case given by Eqs. (2) and (3), we can apply the same method where our problem is described by the equations :

$$\begin{aligned} f &= p_a p_E \sum p^\lambda (p_{d1}^\lambda + p_{d2}^\lambda), \\ g_1 &= p_a p_E \sum p^\lambda p_{d1}^\lambda + (1-p_a)(1+\alpha)p_B - p_{s1}, \\ g_2 &= p_a p_E \sum p^\lambda p_{d2}^\lambda + (1-p_a)(1-\alpha)p_B - p_{s2}, \\ g_c &= p_a p_E \sum p^\lambda p_{d1}^\lambda p_{d2}^\lambda + (1-p_a)(1-\alpha^2)p_B^2 - p_c, \\ \mathcal{L} &= f - \Lambda_1 g_1 - \Lambda_2 g_2 - \Lambda_c g_c, \\ \nabla\mathcal{L} &= 0. \end{aligned} \tag{A10}$$

The optimisation is done taking into account the physical constraints on the attack parameters : all probabilities must be between 0 and 1 and $\text{sign}(p_{d1}^\lambda - p_{d2}^\lambda) \neq f(\lambda)$. The resolution of the system gives us all the extrema of the function $f$. By discarding non-physical solution and taking the highest of the remaining values, we obtain the maximum of Eve's information on the key.